\pdfoutput=1
\documentclass[a4paper]{jpconf}

\usepackage{graphicx,setspace,bm,amsmath}
\usepackage[top=2.5cm,bottom=2.5cm,left=2.5cm,right=2.5cm,footskip=0.75cm]{geometry}

\newcommand\trmi[1]{{\mbox{\scriptsize #1}}} 
\newcommand{\beq}{\begin{equation}} 
\newcommand{\eeq}{\end{equation}} 
\newcommand{\beqa}{\begin{eqnarray}} 
\newcommand{\eeqa}{\end{eqnarray}} 

\begin{document}

{\hspace{9.75truecm}\mbox{BI-TP 2013/23, HIP-2013-24/TH}}

\title{Kurtoses and high order cumulants: Insights from resummed perturbation theory}
\author{Sylvain~Mogliacci}
\address{Faculty of Physics, University of Bielefeld, D-33615 Bielefeld, Germany\\* Department of Physics, P.O.Box 64, FI-00014 University of Helsinki, Finland}
\ead{sylvain@physik.uni-bielefeld.de}

\makeatletter
\renewcommand*{\ps@plain}{\let\@mkboth\@gobbletwo\let\@oddhead\@empty\def\@oddfoot{\reset@font\hfil\thepage}\let\@evenhead\@empty\let\@evenfoot\@oddfoot}
\makeatother
\thispagestyle{plain}
\pagestyle{plain}
\pagenumbering{arabic}

\begin{abstract}
Cumulants of conserved charges provide important information about the physics of the quark-gluon plasma around the phase transition region, as they are by construction sensitive to changes in the degrees of freedom of the system. In this brief proceedings contribution\footnote{This is a slightly modified version of the talk I gave at the workshop Fairness 2013. It is based on two papers~\cite{Paper1,Paper2}. In the paper at hand, I wanted to focus on certain parts only --- more interesting for heavy ion collisions --- especially in the light of recent further developments in resummed perturbative QCD~\cite{HTLptFiniteMUThreeLoop}.}, I report on recent results for two such quantities from two different improved perturbative frameworks, as well as discuss their relevance for heavy ion experiments.
\end{abstract}


\section{Introduction}\label{sec:INTRODUCTION}

The weak coupling expansion of the QCD pressure --- a central quantity for thermodynamics --- shows few signs of apparent convergence, and only at extremely high temperatures (see e.g.~\cite{ConvergenceWeakCouplingExpansion_and_BlaizotDR}). It is therefore crucial to try to improve the situation by attempting to continue the perturbative results as close to the phase transition region as possible. The importance of such work is highlighted by the fact that the quark-gluon plasma created in heavy ion experiments is argued~\cite{SatzHeavyIonCollisions,MullerQCDMatter} to reach a temperature of only a few times the pseudo-critical temperature of the deconfinement transition, $T_\trmi{c}=154\pm 9$MeV~\cite{PeikertTc,BazavovTc}, where the plasma is clearly not extremely weakly coupled.

When the quark chemical potentials $\mu_f$ are turned on to study the phase diagram of the theory (see \cite{RHIC,LHC} for present and \cite{FAIR1,FAIR2,NICA} for future collider experiments), lattice Monte Carlo simulations become inapplicable due to the so-called sign problem of QCD \cite{ForcrandSignPB,GuptaSignPB}. There have been many attempts to find a solution to this problem, but so far the most successful approach amounts to merely Taylor expanding the pressure in powers of $\mu_f/T$, giving access to only moderate densities. The coefficients of such Taylor expansions are, however, related to cumulants of globally conserved quantum numbers in the system, and thus give information about their correlations and fluctuations. Consequently, they turn to be very practical probes of changes occurring in the degrees of freedom of the system across the transition region.

The limitation of lattice methods on the $T$-$\mu$ plane is one of the main reasons, why it is interesting to attack the problem of fluctuations and correlations of conserved charges --- and ultimately the equation of state --- from a resummed perturbation theory point of view. Using such a first principle framework, where there is no sign problem, these quantities can be extended to very large values of the ratio $\mu_f/T$, hence giving information about how the quark-gluon plasma behaves, as it becomes more and more dense.

In this proceedings contribution, I will present recent results for two quantities related to cumulants, obtained using two different frameworks of resummed perturbation theory. I will further comment on the insights that these results can provide for heavy ion physics, as they display highly non-trivial and interesting features.

\section{Statistical QCD and cumulants}\label{sec:RELEVANCEHIC}

Given the Hamiltonian $H_\trmi{QCD}$ of Quantum Chromodynamics, the corresponding partition function $Z_\trmi{QCD}$ can be written in the form
\beq
Z_\trmi{QCD}\left(T,\mu_f;V\right) \equiv \mbox{Tr\ }\exp\left[-\frac{1}{T}\Bigg(H_\trmi{QCD}-\sum_f \mu_f\ Q_f\Bigg)\right] = \mbox{Tr\ }\Big(\rho_\trmi{QCD}\Big)\, \label{PartitionFunction} ,
\eeq
where we have employed the usual notation for thermal averages, $\left\langle\vartheta\right\rangle\equiv\mbox{Tr\ }\left(\vartheta\cdot\rho_\trmi{QCD}\right)/Z_\trmi{QCD}$, that can also be expressed using a path integral representation for $Z_\trmi{QCD}$. Here, $Q_f$ and $\mu_f$ denote respectively conserved charges and the corresponding chemical potentials. In the present work, we will mainly consider the up, down and strange quark conserved charges (with their chemical potentials $\mu_\trmi{u},\ \mu_\trmi{d},\ \mu_\trmi{s}$). However, one can equivalently express the partition function in term of the baryon, electric charge and strangeness conserved numbers (with their chemical potentials $\mu_\trmi{B},\ \mu_\trmi{Q},\ \mu_\trmi{S}$). In addition, the partition function gives access to a host of equilibrium thermodynamic quantities such as the pressure and entropy, following the relations
\beqa
p_\trmi{QCD}&=& \frac{\partial \left(T \log Z_\trmi{QCD}\right)}{\partial V}\ \xrightarrow[V\rightarrow\infty]{ }\ \frac{T}{V}\,\log Z_\trmi{QCD} \, ,\\
S_\trmi{QCD}&=& \frac{\partial \left(T \log Z_\trmi{QCD}\right)}{\partial T} \, .
\eeqa
From equation~(\ref{PartitionFunction}), it is easy to see that the mean and (co)variance of two conserved charges, i.e. $\left\langle Q_f \right\rangle$ and $\left\langle \left(Q_f-\left\langle Q_f \right\rangle\right) \cdot \left(Q_g-\left\langle Q_g \right\rangle\right) \right\rangle$, can be expressed in terms of derivatives with respect to the corresponding chemical potentials. Thus, we can write
\beqa
\left\langle Q_f \right\rangle&=&T\,\frac{\partial}{\partial \mu_f} \log Z_\trmi{QCD} \, , \label{Mean} \\
\left\langle \left(Q_f-\left\langle Q_f \right\rangle\right) \cdot \left(Q_g-\left\langle Q_g \right\rangle\right) \right\rangle&=&T^2\,\frac{\partial^2}{\partial \mu_f \partial \mu_g} \log Z_\trmi{QCD} \, , \label{Covariance}
\eeqa
defining quantities called the first and second order cumulants of the related conserved charges.

Based on the above, it is clear that when considering the covariance of two different charges, such as $Q_\trmi{u}$ and $Q_\trmi{d}$, one is probing the correlation between the two related flavors, while when considering the variance of the same charge, one is probing its fluctuations. The former quantity gives information about possible bound-state survival (see~\cite{BOUNDSTATESURVIVAL1,BOUNDSTATESURVIVAL2} and references therein), while the latter is related to how the system reacts to small increases of density. It also follows that in the infinite volume limit, equations~(\ref{Mean}) and~(\ref{Covariance}) relate the cumulants to the equation of state of the system via derivatives with respect to chemical potentials. These quantities, typically referred to as susceptibilities, are denoted by
\beq
\chi_\trmi{u$_i$\,d$_j$\,s$_k$\, ...}\left(T,\left\{\mu_f\right\}\right) \equiv \frac{\partial^n p_\trmi{QCD}\left(T,\left\{\mu_f\right\}\right)}
{\partial\mu_\trmi{u}^i\, \partial\mu_\trmi{d}^j \, \partial\mu_\trmi{s}^k\, ...} \ ,
\eeq
where $n=i+j+k+...$ and where instead of quark numbers one may also consider any other conserved charge.

The existence and location of a possible critical point on the phase diagram of QCD is often investigated through the behavior of cumulants. Indeed, the variance of the baryon number $\chi_\trmi{B$_2$}\left(T,\mu_\trmi{B}\right)$ is expected to display a critical-like behavior with a sharp peak at the critical point, as it is highly sensitive to changes of density in the medium. This quantity can be estimated via a Taylor expansion containing higher order cumulants at vanishing chemical potential (see e.g.~\cite{SimonTwoFlavors} for discussion and results in the two flavor case). Also, as the convergence radius of the Taylor series of $\chi_\trmi{B$_2$}\left(T,\mu_\trmi{B}\right)$ has been argued to determine the location of a possible critical point (see~\cite{Sayantan} for a recent article), it is clearly interesting to consider higher order cumulants at vanishing chemical potential. For some recent reviews on heavy ion collisions and in particular the use of cumulants in statistical QCD, the reader is referred to~\cite{SatzHeavyIonCollisions,KochCumulant}.

\section{Resummed perturbative QCD}\label{sec:FRAMEWORKS}

\subsection{Dimensional Reduction inspired resummation}

It is well known~\cite{DimensionalReductionPhenomenon1,DimensionalReductionPhenomenon2} that dimensional reduction allows the dynamics of the energy scales of order $gT$ and smaller to be properly accounted for through an effective field theory, which for QCD is called Electrostatic QCD (EQCD). Indeed, by integrating out the hard degrees of freedom, one obtains a three-dimensional SU($N_\trmi{c}$) Yang-Mills theory coupled to an adjoint Higgs field~\cite{BraatenNietoEQCD,KajantieEQCD}. This effective theory turns out to account for all of the infrared divergences typically encountered in naive weak coupling expansions~\cite{LindeIRcatastrophe}. Thus, EQCD not only provides a way out of the so-called infrared catastrophe, but in addition allows for weak coupling calculations to be carried out to high orders in perturbation theory. This, of course, requires a careful matching procedure of the effective theory with full QCD.

Taking advantage of the above fact, one can write the QCD pressure in the form
\beq
p_\trmi{QCD}= p_\trmi{hard}(\mbox{\small$g$}) + T\, p_\trmi{EQCD}(\mbox{\small$m_\trmi{E},g_\trmi{E},\lambda_\trmi{E},\zeta$}) \, ,
\eeq
where the EQCD parameters $m_\trmi{E}(g),g_\trmi{E}(g),\lambda_\trmi{E}(g)$ and $\zeta(g)$ are in turn functions of temperature and the quark chemical potentials. They admit expansions in powers of the four-dimensional gauge coupling $g$. The function $p_\trmi{hard}$ gives the contribution of the hard scale ($\sim T$) only, and is available through a strict loop expansion in QCD, while $p_\trmi{EQCD}$ gives the contribution of the soft ($\sim gT$) and the ultrasoft ($\sim g^2T$) scales, being accessible from the partition function of EQCD.

Typically, when computing a physical quantity through EQCD, one expands the EQCD parameters --- and ultimately the entire result --- in powers of $g$. However, as first suggested in~\cite{ConvergenceWeakCouplingExpansion_and_BlaizotDR} and later successfully applied to the case of the pressure at zero chemical potential in~\cite{MikkoYorkQuarkMassThresholds}, one can alternatively simply consider both $p_\trmi{hard}$ and $p_\trmi{EQCD}$ functions of the EQCD parameters, and not expand them in $g$. This indeed resums certain higher order contributions to the pressure, containing all correct contributions up to and including order $g^6 \log g$~\cite{KeijoMikkoYork,Aleksi1} when expanded in powers of $g$. This procedure leads to a considerable decrease in the theoretical uncertainties of the result through a substantial reduction of its renormalization scale dependence, and thus also to an improvement in its convergence properties.

In the following, we will denote the results originating from the above procedure the `DR ones'. More details about the implementation of the resummation at finite chemical potentials can be found in~\cite{Paper2}, while for general details about dimensional reduction and EQCD, the reader is referred to the original references~\cite{BraatenNietoEQCD,KajantieEQCD} as well as to~\cite{Aleksi1,HartLainePhilipsen,Ipp,AleksiPhDThesis} for a generalization to finite density.

\subsection{Hard-thermal-loop perturbation theory}

Another way of resumming important higher order contributions to thermodynamic quantities is to make use of a variationally improved perturbation theory framework (see~\cite{RefOptKneurAndOthers,RefOptPeterForSPT} and references therein). The basic idea of such a framework is to introduce a physically relevant (usually mass) term that is added and subtracted from the Lagrangian density of the original theory. Treating the added piece with the non-interacting part of the action while adding the subtracted one to the interaction terms of it, one ends up interpolating between the original theory and a theory having dressed propagators and vertices. The procedure is of course such that working consistently to a given order in a perturbative expansion, one in the end recovers results corresponding to the correct original theory.

In a gauge field theory such as QCD, the situation is somewhat more complicated, as one cannot simply add and subtract a local mass term for the gauge fields without breaking the gauge invariance. Instead, one can make use of the well-known hard-thermal-loop effective action, first derived in~\cite{FrenkelTaylorHTL,BraatenPisarskiHTL}, which plays the role of a non-local (and hence momentum dependent) mass term. With the QCD pressure, this approach was first applied more than a decade ago in~\cite{JensMikeBraatenFirstHTLpt,JensMikeBraatenFirstHTLptbis,JensMikeFirstTwoLoop}, and is known as Hard-Thermal-Loop perturbation theory (HTLpt).

In short, HTLpt amounts to reorganizing the Lagrangian density of QCD as
\beq
{\cal L_{\rm HTLpt}}=\Big[{\cal L}_{\rm QCD}+(1-\delta)\ {\cal L}_{\rm HTL}\Big]\Big|_{g\rightarrow\sqrt{\delta}g}+\Delta{\cal L}_{\rm HTL} \, ,
\eeq
where ${\cal L}_{\rm HTL}$ is a gauge invariant HTL improvement term, given by the HTL effective action, and $\delta$ a formal expansion parameter set to one after having Taylor expanded the path integral up to some order in it. Finally, $\Delta{\cal L}_{\rm HTL}$ is a counter term necessary to cancel further ultraviolet divergences introduced by the resulting reorganization of the perturbative series. Essentially, this reorganization amounts to shifting the ground state of the expansion from an ideal gas of massless particles to an ideal gas of massive quasiparticles~\cite{HTLptFiniteMUThreeLoop,NanHTLpt1,NanHTLpt2}.

More details about the implementation of the HTLpt procedure at finite chemical potentials can be found in~\cite{Paper1,Paper2} as well as in~\cite{HTLptFiniteMUThreeLoop,HTLptFiniteMUTwoLoop1,HTLptFiniteMUTwoLoop2}. For related developments, see also~\cite{BlaizotHTLResummation1,BlaizotHTLResummation2}.

\subsection{Technical details}

Working at four- and one-loop levels in the DR and HTLpt schemes, respectively, we must assign values to various parameters within them. As is customary in perturbative thermal QCD calculations, the renormalization scale is chosen to be of the order of $2\pi T$, and then varied by a factor of two around this central value in order to estimate the sensitivity of the results with respect to this choice. Following~\cite{FAC}, we choose to apply the Fastest Apparent Convergence (FAC) scheme to the next-to-leading order expression of the gauge coupling of EQCD. Thereby, we obtain $\bar{\Lambda}_{\rm central}=1.445\times 2\pi T$ and $1.291\times 2\pi T$ for the central values with two and three flavors, respectively~\cite{Paper1,Paper2}. Regarding the running of the coupling, we use a two-loop expression in the DR case~\cite{MikkoYorkTwoLoopG} and one-loop running in the HTLpt result. Finally, to fix the value of $\Lambda_{\rm QCD}$ (in the $\overline{\trmi{MS}}$ scheme), we use the recent lattice result $\alpha_\trmi{s}(1.5\;{\rm GeV})=0.326$~\cite{AlphaS}, demanding that our one- and two-loop couplings agree with it for $\bar{\Lambda}=1.5$ GeV. This respectively gives $\Lambda_{\rm QCD}=176$ MeV and 283 MeV for $N_\trmi{f}=3$, as well as $\Lambda_{\rm QCD}=204$ MeV and 324 MeV for $N_\trmi{f}=2$. These values are then varied by $\pm$ 30 MeV.

In the plots appearing in the following section, the bands corresponding to our perturbative results are composed by varying all of the parameters mentioned here in their respective ranges.

\section{Results}\label{sec:RESULTS}

\subsection{Kurtoses}

\begin{figure}[!ht]\centering\includegraphics[scale=0.30]{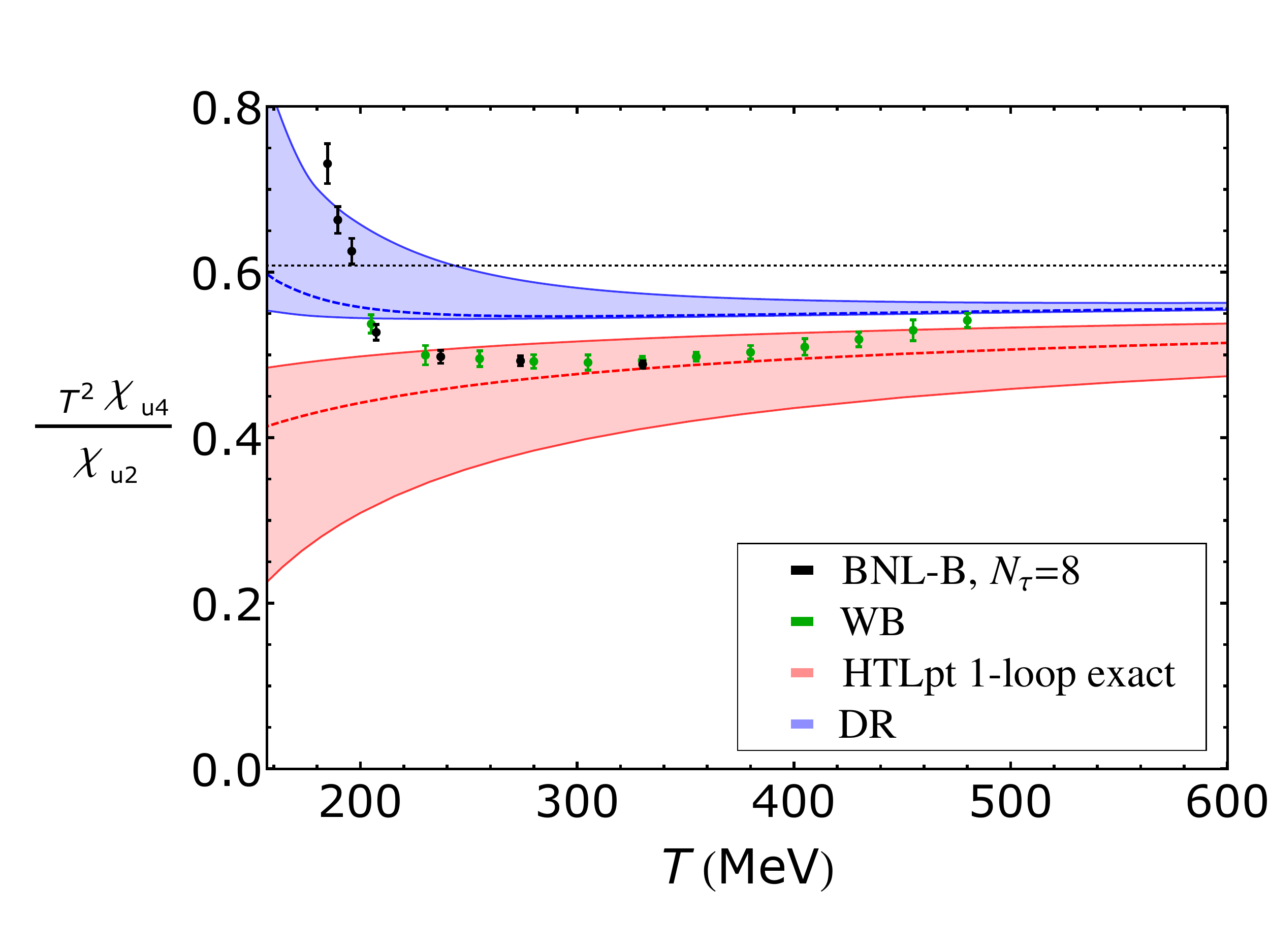}\!\!\!\!\includegraphics[scale=0.30]{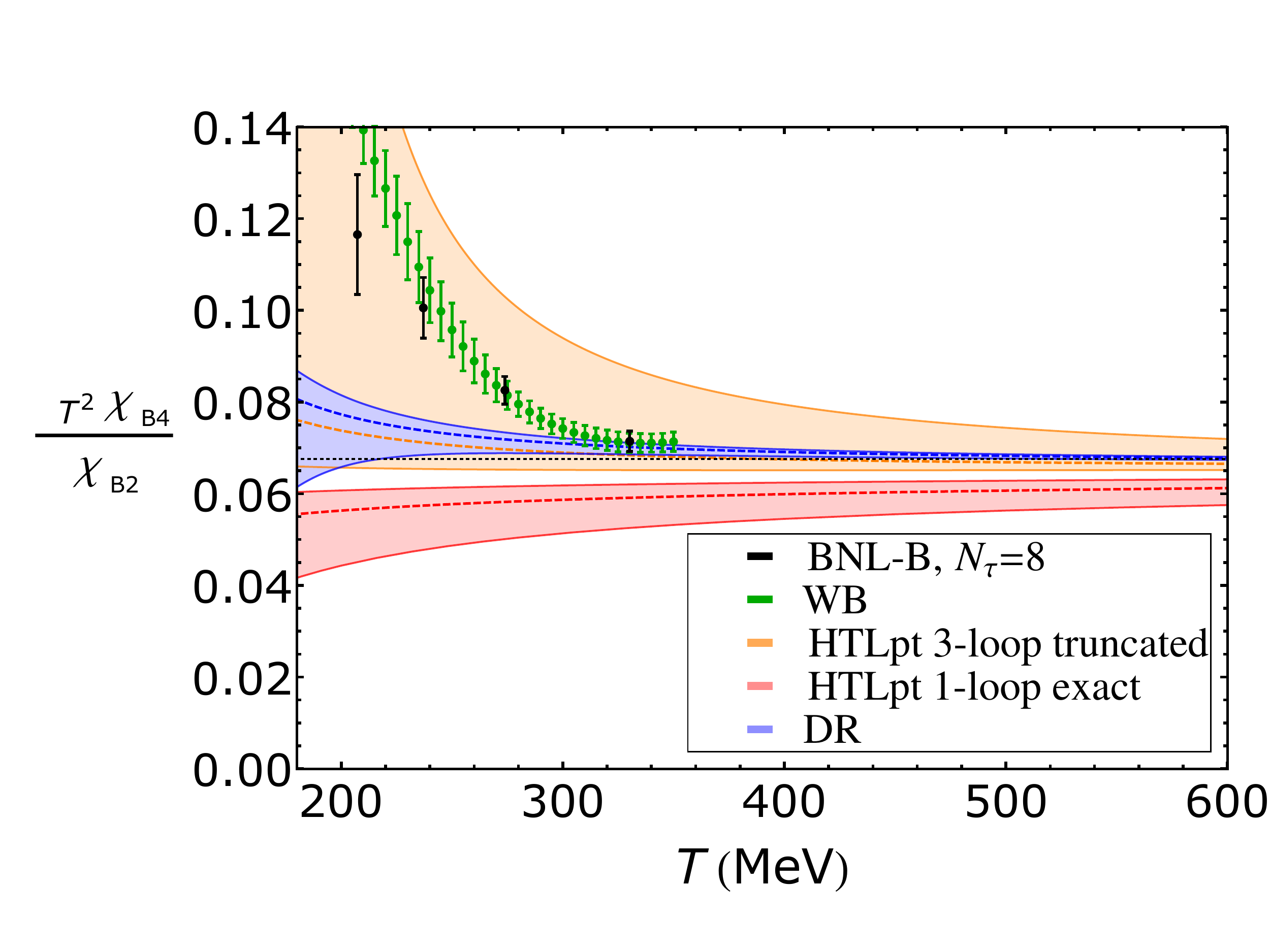}
\caption{Ratios of cumulants related to the light quark (left) and baryon (right) numbers, with lattice data taken from~\cite{RatioBNLB1,RatioBNLB2} (BNL-B) and~\cite{RatioWuppBudLight,RatioWuppBudBaryon} (WB). The three-loop HTLpt result is obtained from the corresponding cumulants of~\cite{HTLptFiniteMUThreeLoop}. The dashed curves inside the bands correspond to the central values of the renormalization and QCD scales, while the black dashed (straight) lines denote the Stefan Boltzmann limits.\label{ratios}}
\end{figure}

Roughly speaking, the kurtosis is a measure of how strongly peaked a given quantity is, and is defined as the ratio of the corresponding fourth and second order cumulants, the latter one squared. In our figure~\ref{ratios}, we display results for this quantity for the (light) quark and baryon numbers; we further multiply the results by the second order cumulants, so that in effect we simply plot ratios of the fourth and second order cumulants. Such a quantity can be used as a measure of how fast a phase transition is; see e.g.~the discussion of \cite{StephanovKurtosis} with regards to the behavior of the kurtosis relevant to the order parameter of the deconfinement phase transition near the critical point.

In figure~\ref{ratios} (left), we plot our DR and HTLpt results together with lattice data from~\cite{RatioBNLB1,RatioBNLB2,RatioWuppBudLight}, which at temperatures around $T=$ 300 -- 400 MeV seem to agree nicely with the one-loop HTLpt band, but soon start approaching the DR prediction. It should be noted that the latter reproduces the overall trend of the lattice data better. On the right figure, we see similar trends in both predictions, with the difference that the DR one seems to converge to the Stefan Boltzmann limit much faster than in the case of the light quark kurtosis. This trend is seen to be in accordance with the displayed lattice data from~\cite{RatioBNLB2,RatioWuppBudBaryon}, and agrees with the expectation that the medium is much less sensitive to the hadronic degrees of freedom in this range of temperatures. Moreover, in figure~\ref{ratios} (right), we display the corresponding three-loop HTLpt prediction obtained using the cumulants computed in~\cite{HTLptFiniteMUThreeLoop}. Although the band is relatively large at low temperature, this result is also seen to reproduce the qualitative trend of the lattice data, furthermore agreeing quite well with the DR one. From the one-loop to the three-loop HTLpt prediction, we notice a reasonably good convergence of the series for this quantity.

\subsection{Sixth order cumulants}

\begin{figure}[!ht]\centering\includegraphics[scale=0.30]{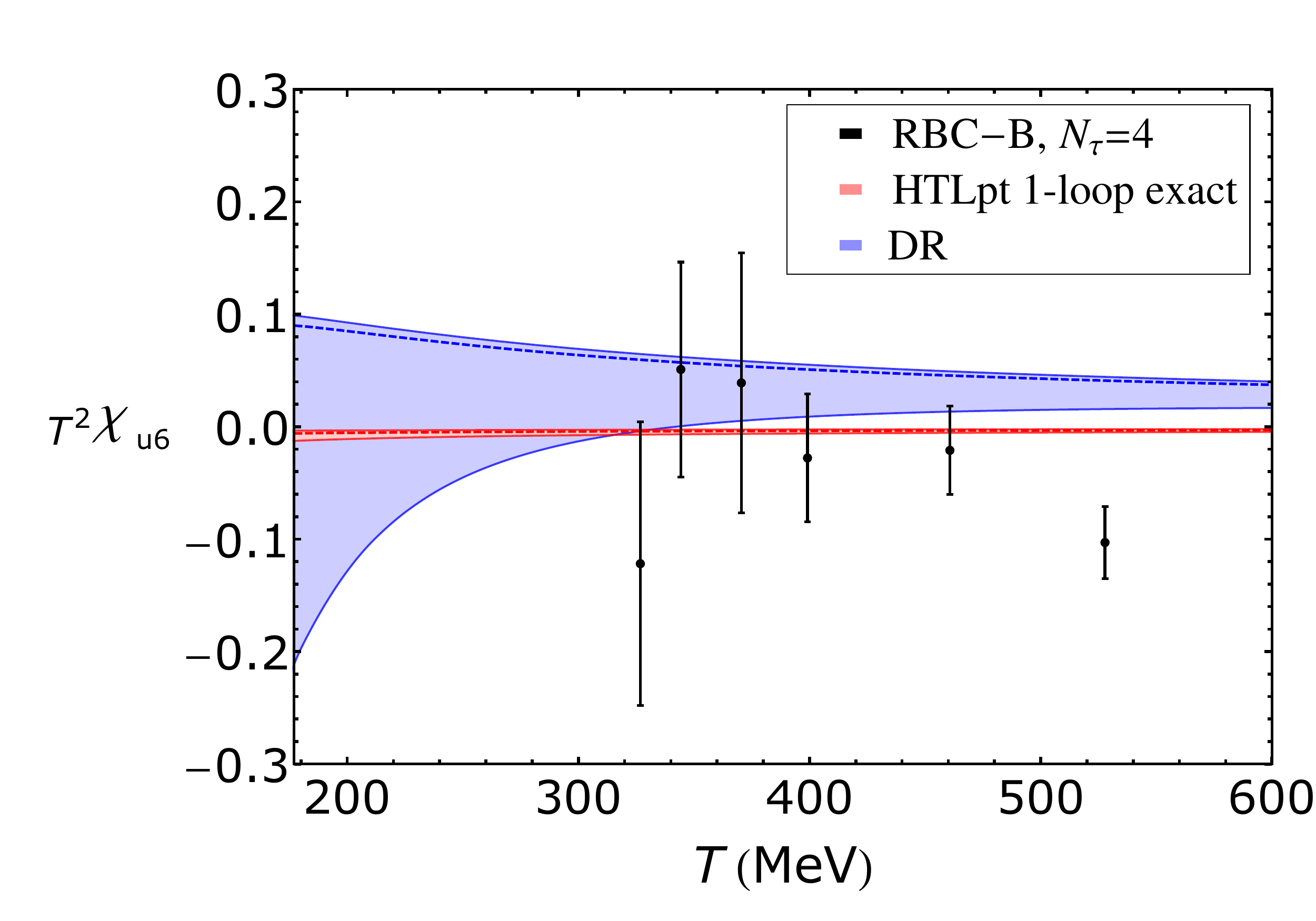}\!\!\!\!\includegraphics[scale=0.30]{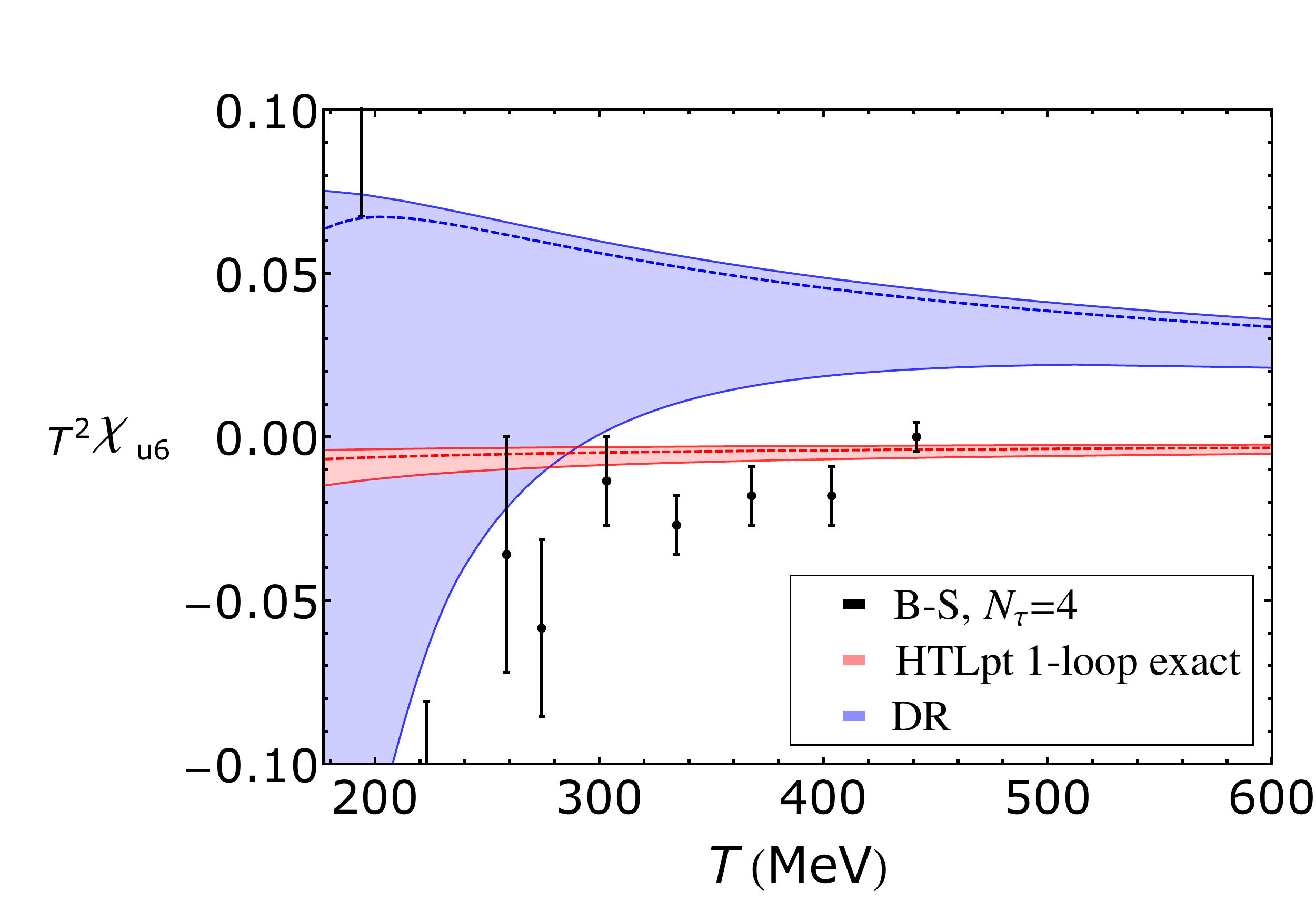}
\caption{The sixth order cumulant for the light quark number, evaluated for $N_\trmi{f}=3$ (left) and $N_\trmi{f}=2$ (right), with lattice data from~\cite{Peterchi6} (left) and~\cite{SimonTwoFlavors} (right). The right figure has been taken from~\cite{Paper2}. The dashed curves inside the bands correspond to the central values of the renormalization and QCD scales.\label{chi6Nf3and2}}
\end{figure}

In figure~\ref{chi6Nf3and2}, we next display sixth order cumulants for the light quark number in the cases of three (left) and two (right) quark flavors. Such higher order cumulants are expected to be very sensitive probes of the hadronic freeze-out~\cite{KarschFluctu}. For these quantities, the two first orders of perturbation theory vanish, and the weak coupling expansions only start at order $g^3$, rendering the properties of the weak coupling series much poorer than for the lower order cumulants. Indeed, we see from both the left and right figures that our DR predictions are positive for all displayed temperatures, while the HTLpt results are consistently negative. The latter apparently agree better with the (unfortunately not continuum extrapolated) lattice data of~\cite{SimonTwoFlavors,Peterchi6}. However, it should be recalled our HTLpt predictions are of one-loop order, and based on the very good agreement between our DR results and the recent three-loop HTLpt calculations of~\cite{HTLptFiniteMUThreeLoop} (see~\cite{Paper2} for a detailed comparison), it is probable that the forthcoming three-loop HTLpt prediction of~\cite{THREELOOPHTLptforth} for the sixth order cumulant will also turn out to be positive.

It is interesting to note that the leading plasmon contribution to the weak coupling expansion of the sixth order cumulant is negative, while it can be seen by truncating our DR results at various orders that the sign of the result is turned positive by the higher order contributions. Noting that lattice studies seem to favor negative values for sixth order cumulants~\cite{SimonTwoFlavors,KarschFluctu,Peterchi6} above the transition region, we suspect that including further perturbative orders to the current DR result, it will eventually change in the temperature range considered. This will, however, likely not be verified in the near future, as the technical challenges involved in e.g.~a full five-loop determination of the quantity are formidable.

\section{Conclusion}\label{sec:CONLUSION}

In this proceedings contribution, which is based on the recent resummed perturbation theory investigations~\cite{Paper1,Paper2}, we have reviewed predictions for two quantities highly relevant for the physics of heavy ion collisions.

First, considering kurtoses corresponding to the light quark and baryon number operators for three flavor QCD, we saw how resummed perturbation theory provides a very successful description of lattice data down to 2 -- 3 $T_\trmi{c}$. This is clearly an indication that a weakly interacting quasiparticles picture is well suited for the fermionic sector of the theory, even close to the transition region. Also, information on the relevant degrees of freedom was highlighted by the qualitative difference of fluctuations between light quark and baryon conserved charges.

Next, we looked at sixth order cumulants in the cases of two and three flavor QCD. Here, we encountered quantities with which resummed perturbation theory has a much harder time, which could be understood from the fact that the weak coupling expansions only start at a relatively high order.

Quantities like the ones considered here are important for understanding the phase transition leading to the formation of the quark-gluon plasma, and also give crucial information about changes in the degrees of freedom of the system right above the transition. I therefore hope that these results, and more generally resummed perturbation theory as a complementary first principle tool, will turn out to be of practical use in the present and future analysis of heavy ion collision data.

\ack
I am grateful to P\'eter Petreczky and Aleksi Vuorinen for valuable discussions, helpful in the preparation of this manuscript. I would also like to thank Nan Su for introducing this workshop to me, as well as the Helsinki Institute of Physics for its hospitality, where this proceedings contribution was written. This work has been supported by the mobility grant of the Bielefeld Graduate School in Theoretical Sciences.



\section*{References}
\begin{thebibliography}{99}


\bibitem{Paper1}
Andersen J~O~, Mogliacci S~, Su N~and Vuorinen A~;
2013;~Quark number susceptibilities from resummed perturbation theory;
~{\it{Phys.\ Rev.}} D {\bf{87}} 074003 ({\it{Preprint~}} arXiv:1210.0912)

\bibitem{Paper2}
Mogliacci S~, Andersen J~O~, Strickland M~, Su N~and Vuorinen A~;
2013;~Equation of State of hot and dense QCD: Resummed perturbation theory confronts lattice data;
~{\it{Preprint~}} arXiv:1307.8098

\bibitem{HTLptFiniteMUThreeLoop}
Haque N~, Andersen J~O~, Mustafa M~G~, Strickland M~and Su N~;
2013;~Three-loop HTLpt Pressure and Susceptibilities at Finite Temperature and Density;
~{\it{Preprint~}} arXiv:1309.3968


\bibitem{ConvergenceWeakCouplingExpansion_and_BlaizotDR}
Blaizot J~P~, Iancu E~and Rebhan A~;
2003;~On the apparent convergence of perturbative QCD at high temperature;
~{\it{Phys.\ Rev.}} D {\bf{68}} 025011 ({\it{Preprint~}} hep-ph/0303045)

\bibitem{SatzHeavyIonCollisions}
Satz H~;
2013;~Probing the States of Matter in QCD;
~{\it{Int.\ J.\ Mod.\ Phys.}} A {\bf{28}} 1330043 ({\it{Preprint~}} arXiv:1310.1209)

\bibitem{MullerQCDMatter}
M\"{u}ller B~;
2013;~Investigation of Hot QCD Matter: Theoretical Aspects;
~{\it{Preprint~}} arXiv:1309.7616

\bibitem{PeikertTc}
Karsch F~, Laermann E~and Peikert A~;
2001;~Quark mass and flavor dependence of the QCD phase transition;
~{\it{Nucl.\ Phys.}} B {\bf{605}} 579 ({\it{Preprint~}} hep-lat/0012023)

\bibitem{BazavovTc}
Bazavov A~{\it et al.}~;
2012;~The chiral and deconfinement aspects of the QCD transition;
~{\it{Phys.\ Rev.}} D {\bf{85}} 054503 ({\it{Preprint~}} arXiv:1111.1710)

\bibitem{RHIC}
Tannenbaum M~J~;
2012;~Highlights from BNL-RHIC;
~{\it{Preprint~}} arXiv:1201.5900

\bibitem{LHC}
M\"{u}ller B~, Schukraft J~and Wyslouch B~;
2012;~First Results from Pb+Pb collisions at the LHC;
~{\it{Ann.\ Rev.\ Nucl.\ Part.\ Sci.}} {\bf{62}} 361 ({\it{Preprint~}} arXiv:1202.3233)

\bibitem{FAIR1}
Heuser J~M~for the CBM collaboration;
2013;~The compressed baryonic matter experiment at FAIR;
~{\it{Nucl.\ Phys.}} A {\bf{904-905}} 941c

\bibitem{FAIR2}
Senger P~, Galatyuk T~, Kresan D~, Kiseleva A~and Kryshen E~;
2006;~CBM at FAIR;
~{\it{PoS}} {\bf{CPOD2006}} 018

\bibitem{NICA}
Kekelidze V~, Kovalenko A~, Lednicky R~, Matveev V~, Meshkov I~, Sorin A~and Trubnikov G~;
2013;~Project NICA at JINR;
~{\it{Nucl.\ Phys.}} A {\bf{904-905}} 945c

\bibitem{ForcrandSignPB}
de~Forcrand P~;
2009;~Simulating QCD at finite density;
~{\it{PoS}} {\bf{LAT2009}} 010  ({\it{Preprint~}} arXiv:1005.0539)

\bibitem{GuptaSignPB}
Gupta S~;
2010;~QCD at finite density;
~{\it{PoS}} {\bf{LAT2010}} 007  ({\it{Preprint~}} arXiv:1101.0109)


\bibitem{BOUNDSTATESURVIVAL1}
Krieg S~{\it{et al.}} for the Wuppertal-Budapest collaboration;
2013;~Fluctuations of conserved charges at finite temperature from lattice QCD;
{\it{J.\ Phys.\ Conf.\ Ser.}} {\bf{432}} 012012

\bibitem{BOUNDSTATESURVIVAL2}
Ratti C~, Bellwied R~, Cristoforetti M~and Barbaro M~;
2012;~Are there hadronic bound states above the QCD transition temperature?;
{\it{Phys.\ Rev.}} D {\bf{85}} 014004 ({\it{Preprint~}} arXiv:1109.6243)

\bibitem{SimonTwoFlavors}
Allton C~R~, Doring M~, Ejiri S~, Hands S~J~, Kaczmarek O~, Karsch F~, Laermann E~and Redlich K~;
2005;~Thermodynamics of two flavor QCD to sixth order in quark chemical potential;
~{\it{Phys.\ Rev.}} D {\bf{71}} 054508 ({\it{Preprint~}} hep-lat/0501030)

\bibitem{Sayantan}
Gavai R~V~and Sharma S~;
2012;~A faster method of computation of lattice quark number susceptibilities;
~{\it{Phys.\ Rev.}} D {\bf{85}} 054508 ({\it{Preprint~}} arXiv:1112.5428)

\bibitem{KochCumulant}
Koch V~;
2008;~Hadronic Fluctuations and Correlations;
~{\it{Preprint~}} arXiv:0810.2520


\bibitem{DimensionalReductionPhenomenon1}
Appelquist T~and Pisarski R~D~;
1981;~High-Temperature Yang-Mills Theories and Three-Dimensional Quantum Chromodynamics;
~{\it{Phys.\ Rev.}} D {\bf{23}} 2305

\bibitem{DimensionalReductionPhenomenon2}
Nadkarni S~;
1983;~Dimensional reduction in finite-temperature quantum chromodynamics;
~{\it{Phys.\ Rev.}} D {\bf{27}} 917

\bibitem{BraatenNietoEQCD}
Braaten E~and Nieto A~;
1995;~Effective field theory approach to high temperature thermodynamics;
~{\it{Phys.\ Rev.}} D {\bf{51}} 6990 ({\it{Preprint~}} hep-ph/9501375)

\bibitem{KajantieEQCD}
Kajantie K~, Laine M~, Rummukainen K~and Shaposhnikov M~E~;
1996;~Generic rules for high temperature dimensional reduction and their application to the standard model;
~{\it{Nucl.\ Phys.}} B {\bf{458}} 90 ({\it{Preprint~}} hep-ph/9508379)

\bibitem{LindeIRcatastrophe}
Linde A~D~;
1980;~Infrared Problem in Thermodynamics of the Yang-Mills Gas;
~{\it{Phys.\ Lett.}} B {\bf{96}} 289

\bibitem{MikkoYorkQuarkMassThresholds}
Laine M~and Schr\"{o}der Y~;
2006;~Quark mass thresholds in QCD thermodynamics;
~{\it{Phys.\ Rev.}} D {\bf{73}} 085009 ({\it{Preprint~}} hep-ph/0603048)

\bibitem{KeijoMikkoYork}
Kajantie K~, Laine M~, Rummukainen K~and Schr\"{o}der Y~;
2003;~The Pressure of hot QCD up to $g^6 \ln(1/g)$;
~{\it{Phys.\ Rev.}} D {\bf{67}} 105008 ({\it{Preprint~}} hep-ph/0211321)

\bibitem{Aleksi1}
Vuorinen A~;
2003;~The Pressure of QCD at finite temperatures and chemical potentials;
~{\it{Phys.\ Rev.}} D {\bf{68}} 054017 ({\it{Preprint~}} hep-ph/0305183)

\bibitem{HartLainePhilipsen}
Hart A~, Laine M~and Philipsen O~;
2000;~Static correlation lengths in QCD at high temperatures and finite densities;
~{\it{Nucl.\ Phys.}} B {\bf{586}} 443 ({\it{Preprint~}} hep-ph/0004060)

\bibitem{Ipp}
Ipp A~, Kajantie K~, Rebhan A~and Vuorinen A~;
2006;~The Pressure of deconfined QCD for all temperatures and quark chemical potentials;
~{\it{Phys.\ Rev.}} D {\bf{74}} 045016 ({\it{Preprint~}} hep-ph/0604060)

\bibitem{AleksiPhDThesis}
Vuorinen A~;
2004;~The Pressure of QCD at finite temperature and quark number density -- PhD thesis;
~{\it{Preprint~}} hep-ph/0402242

\bibitem{RefOptKneurAndOthers}
Kneur J~-L~and Neveu A~;
2013;~$\alpha_S$ from $F_\pi$ and Renormalization Group Optimized Perturbation;
~{\it{Phys.\ Rev.}} D {\bf{88}} 074025 ({\it{Preprint~}} arXiv:1305.6910)

\bibitem{RefOptPeterForSPT}
Karsch F~, Patk\'{o}s A~and Petreczky P~;
1997;~Screened perturbation theory;
~{\it{Phys.\ Lett.}} B {\bf{401}} 69 ({\it{Preprint~}} hep-ph/9702376)

\bibitem{FrenkelTaylorHTL}
Frenkel J~and Taylor J~C~;
1990;~High Temperature Limit of Thermal QCD;
~{\it{Nucl.\ Phys.}} B {\bf{334}} 199

\bibitem{BraatenPisarskiHTL}
Braaten E~and Pisarski R~D~;
1990;~Soft Amplitudes in Hot Gauge Theories: A General Analysis;
~{\it{Nucl.\ Phys.}} B {\bf{337}} 569

\bibitem{JensMikeBraatenFirstHTLpt}
Andersen J~O~, Braaten E~and Strickland M~;
2000;~Hard thermal loop resummation of the thermodynamics of a hot gluon plasma
~{\it{Phys.\ Rev.}} D {\bf{61}} 014017 ({\it{Preprint~}} hep-ph/9905337)

\bibitem{JensMikeBraatenFirstHTLptbis}
Andersen J~O~, Braaten E~and Strickland M~;
2000;~Hard thermal loop resummation of the free energy of a hot quark-gluon plasma
~{\it{Phys.\ Rev.}} D {\bf{61}} 074016 ({\it{Preprint~}} hep-ph/9908323)

\bibitem{JensMikeFirstTwoLoop}
Andersen J~O~, Braaten E~, Petitgirard E~and Strickland M~;
2002;~HTL perturbation theory to two loops;
~{\it{Phys.\ Rev.}} D {\bf{66}} 085016 ({\it{Preprint~}} hep-ph/0205085)

\bibitem{NanHTLpt1}
Andersen J~O~, Strickland M~and Su N~;
2010;~Three-loop HTL gluon thermodynamics at intermediate coupling;
~{\it{J.\ High\ Energy\ Phys.}} {\bf{1008}} 113 ({\it{Preprint~}} arXiv:1005.1603)

\bibitem{NanHTLpt2}
Andersen J~O~, Leganger L~E~, Strickland M~and Su N~;
2011;~Three-loop HTL QCD thermodynamics;
~{\it{J.\ High\ Energy\ Phys.}} {\bf{1108}} 053 ({\it{Preprint~}} arXiv:1103.2528)

\bibitem{HTLptFiniteMUTwoLoop1}
Haque N~, Mustafa M~G~and Strickland M~;
2013;~Two-loop HTL pressure at finite temperature and chemical potential;
~{\it{Phys.\ Rev.}} D {\bf{87}} 105007 ({\it{Preprint~}} arXiv:1212.1797)

\bibitem{HTLptFiniteMUTwoLoop2}
Haque N~, Mustafa M~G~and Strickland M~;
2013;~Quark Number Susceptibilities from Two-Loop Hard Thermal Loop Perturbation Theory;
~{\it{J.\ High\ Energy\ Phys.}} {\bf{1307}} 184 ({\it{Preprint~}} arXiv:1302.3228)

\bibitem{BlaizotHTLResummation1}
Blaizot J~P~, Iancu E~and Rebhan A~;
2001;~Approximately selfconsistent resummations for the thermodynamics of the quark gluon plasma. 1. Entropy and density
~{\it{Phys.\ Rev.}} D {\bf{63}} 065003 ({\it{Preprint~}} hep-ph/0005003)

\bibitem{BlaizotHTLResummation2}
Blaizot J~P~, Iancu E~and Rebhan A~;
2001;~Quark number susceptibilities from HTL resummed thermodynamics
~{\it{Phys.\ Lett.}} B {\bf{523}} 143 ({\it{Preprint~}} hep-ph/0110369)

\bibitem{FAC}
Kajantie K~, Laine M~, Rummukainen K~, Shaposhnikov M~E~;
1997;~3-D SU(N) + adjoint Higgs theory and finite temperature QCD;
~{\it{Nucl.\ Phys.}} B {\bf{503}} 357 ({\it{Preprint~}} hep-ph/9704416)

\bibitem{MikkoYorkTwoLoopG}
Laine M~and Schr\"{o}der Y~;
2005;~Two-loop QCD gauge coupling at high temperatures;
~{\it{J.\ High\ Energy\ Phys.}} {\bf{0503}} 067 ({\it{Preprint~}} hep-ph/0503061)

\bibitem{AlphaS}
Bazavov A~, Brambilla N~, Garcia X~, Petreczky P~, Soto J~and Vairo A~;
2012;~Determination of $\alpha_s$ from the QCD static energy;
~{\it{Phys.\ Rev.}} D {\bf{86}} 11403 ({\it{Preprint~}} arXiv:1205.6155)


\bibitem{StephanovKurtosis}
Stephanov M~A~;
2011;~On the sign of kurtosis near the QCD critical point;
~{\it{Phys.\ Rev.\ Lett.}} {\bf{107}} 052301 ({\it{Preprint~}} arXiv:1104.1627)

\bibitem{RatioBNLB1}
Schmidt C~for the BNL-Bielefeld collaboration;
2013;~Baryon number and charge fluctuations from lattice QCD;
~{\it{Nucl.\ Phys.}} A {\bf{904-905}} 865c ({\it{Preprint~}} arXiv:1212.4278)

\bibitem{RatioBNLB2}
Schmidt C~;
2013;~QCD bulk thermodynamics and conserved charge fluctuations with HISQ fermions;
~{\it{J.\ Phys.\ Conf.\ Ser.}} {\bf{432}} 012013 ({\it{Preprint~}} arXiv:1212.4283)

\bibitem{RatioWuppBudLight}
Bors\'{a}nyi S~;
2013;~Thermodynamics of the QCD transition from lattice;
~{\it{Nucl.\ Phys.}} A {\bf{904-905}} 270c ({\it{Preprint~}} arXiv:1210.6901)

\bibitem{RatioWuppBudBaryon}
S.~Bors\'{a}nyi, Z.~Fodor, S.~D.~Katz, S.~Krieg, C.~Ratti and K.~K.~Szabo;
2013;~Freeze-out parameters: lattice meets experiment;
~{\it{Phys.\ Rev.\ Lett.}} {\bf{111}} 062005 ({\it{Preprint~}} arXiv:1305.5161)

\bibitem{KarschFluctu}
Friman B~, Karsch F~, Redlich K~and Skokov V~;
2011;~Fluctuations as probe of the QCD phase transition and freeze-out in heavy ion collisions at LHC and RHIC;
~{\it{Eur.\ Phys.\ J.}} C {\bf{71}} 1694 ({\it{Preprint~}} arXiv:1103.3511)

\bibitem{Peterchi6}
Petreczky P~{\it{et al.}} for the RBC-Bielefeld collaboration;
2009;~Quark number fluctuations at high temperatures;
~{\it{PoS}} {\bf{LAT2009}} 159  ({\it{Preprint~}} arXiv:0911.0196)

\bibitem{THREELOOPHTLptforth}
Haque N~, Andersen J~O~, Mustafa M~G~, Strickland M~and Su N~;
~in preparation

\end{thebibliography}
\end{document}